# An Evolutionary Note on Smart City Development in China[*]

Ruizhi Liao[1][2][‡], Liping Chen[3]

[1]*School of Humanities and Social Science, The Chinese University of Hong Kong, Shenzhen, China*
[2]*Shenzhen Key Laboratory of IoT Intelligent Systems and Wireless Network Technology, China*
[3]*Enterprise & Technology Security, IBM Ireland*
E-mail: rzliao@cuhk.edu.cn; lipingch@ie.ibm.com

**Abstract:** In response to challenges posed by urbanization, David Bollier from the University of Southern California raised a new idea for city planning: a comprehensive network and applications of information technologies. IBM later echoed the idea and initiated its Smart Planet vision in 2008. After that, the smart city concept was quickly adopted by major cities throughout the world, and it has gradually evolved into a strategic choice by ambitious cities. This paper looks into the smart city trend by reviewing how the concept of smart city was proposed and what the essence of a smart city is. More specifically, the driving forces of the smart city development in China are investigated, and the key differences of smart cities between China and other countries are summarized. Finally, four big challenges to build future smart cities are discussed.

## 1 Introduction

Urbanization is an important indicator of a country's degree of modernization. According to the World Urbanization Report by the United Nations Department of Economic and Social Affairs, more than 55% of the world's population lives in cities, with this figure projected to rise to 70% by 2050 (United Nations, 2018). China's urban population accounts for 60% of the national total in 2020, and is expected to reach a very ambitious rate, 80%, by 2050 (Deloitte, 2018).

More people living in cities means more contests over limited urban resources, such as water, housing, transportation, education, health care and so on. In order to cope with the challenges brought by the urbanization trend, David Bollier from the University of Southern California raised a new idea for city planning: a comprehensive network and applications of information technologies (Bollier, 1998).

IBM later echoed the idea and initiated its Smart Planet vision in 2008 (Palmisano, 2008). At that time, it was a brand new concept to empower and transform urban infrastructure, services and management. Quickly afterwards, the smart city concept has been adopted by major cities throughout the world, and it has gradually evolved into a strategic choice by ambitious cities - aiming to compete for urban development, and easy-to-use but efficient functionalities.

This paper discusses the essence of a smart city, and investigates three driving forces of the smart city development in China. This paper also summarizes the key features of smart cities in China and other countries. Finally, this paper raises four big challenges that deserve careful thoughts on building future smart cities.

## 2 The Essence of a Smart City

Let us first look at what a city means before we dive into smart cities. From the societal perspective, a city is a place in which people live and work; from the technological perspective, it is a collection of functional units (such as transportation, education, medical care and utility networks); from the economic perspective, it is where population, industry and commerce converge, and it is also where job opportunities, innovation and wealth are created. The accumulation of knowledge and wealth improves the city's productivity, which in turn further promotes urban development. In this regard, the development of cities is an important driving force for the progress of human society.

---

[‡] Corresponding author
[*] This work was supported in part by NSFC (61902332) and Shenzhen STIC (JCYJ20170818103636337, JCYJ20180508162604311).

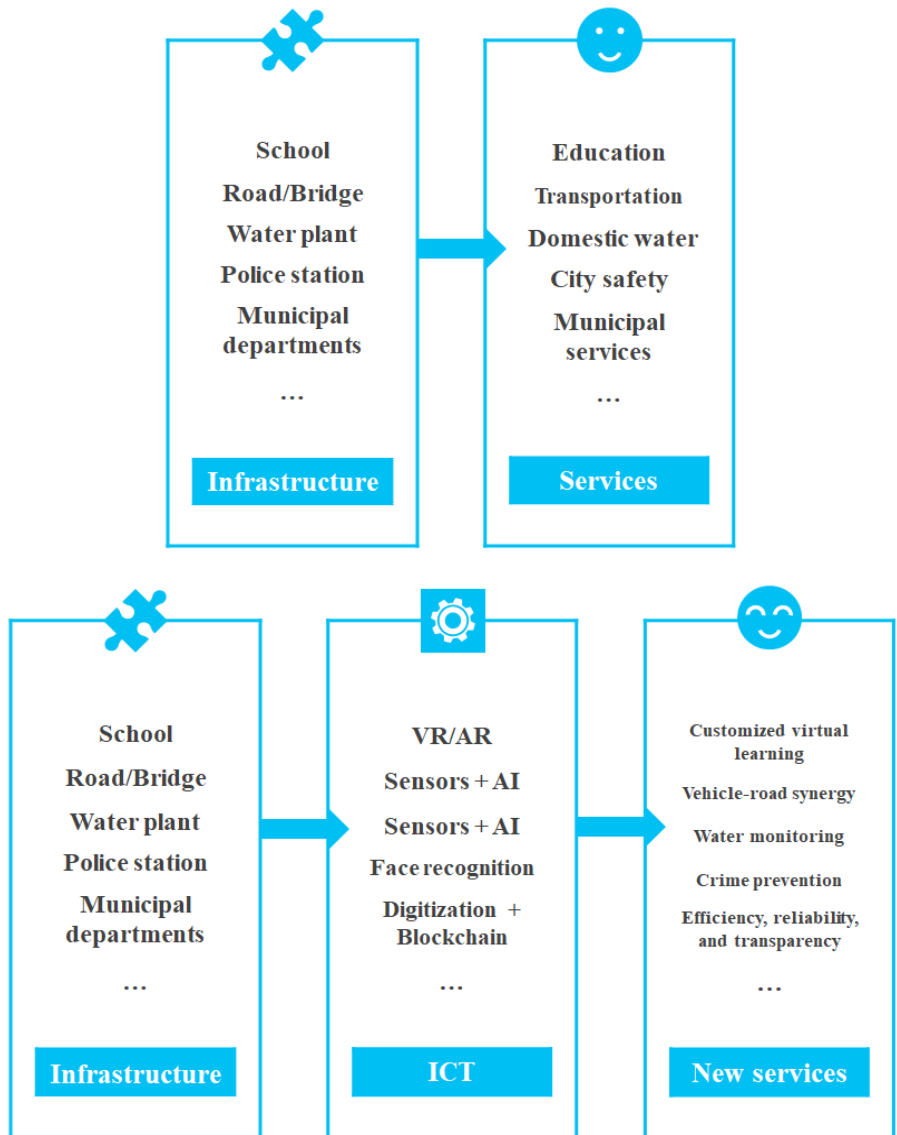

**Fig. 1  The essence of a traditional city vs a smart city**

Then, what is a smart city? Although there is no universal definition for smart city, it can be generally understood as a city that applies information and communication technologies, to empower its infra-structure and functional units in ways that improve its services, operations and management, and also to effectively utilize its resources to achieve sustainable urban development (Albino, et al., 2015). In this regard, the smart city is more related to the technological factor of a city outlined above, and the application of new technologies is indeed one of the key enabling factors for smart city. However, other factors also play important roles in helping us understand what a smart city really is. For example, first, a city's economy affects the construction of its functional units and infrastructure (the economic factor). Second, the priority to construct a functional unit involves negotiation and compromise among stakeholders (the political factor). Third, the introduction of information and communication technologies to a functional unit al-ters or replaces the existing system, and affects the life and work patterns, to which individuals have been accustomed. This may result in objections or unexpected negative effects (the production relation or social factor). Thus, the designing of a smart city must therefore consider not only technological, but also economic, political, societal factors, and their interactions.

So, what is the essence of a smart city? As known, a city's infrastructure is composed of functional units. It is these units and the services they provide that enable citizens to live and work in a particular urban area. The essence of a city thus has two components: the infrastructure, and the services related to it, as shown in the upper part of Fig. 1. The essence of a smart city is to embed information and communication technologies in the gap between infrastructure and services, and use them to optimize the urban operations and daily lives of citizens, as depicted in the lower part of Fig. 1. For example, by incorporating technologies like augmented reality (AR), virtual reality (VR) or mixed reality (MR) into traditional education, we can now provide students with an immersive learning environment with benefits such as personalized education, better student engagement and better content understanding (Yagol, et al., 2018). In another example, by employing sensors at water plants and home water meters, we can now monitor water quality, identify water leakage, and suggest water saving plans based on the water usage statistics (Chen & Han, 2018).

## 3 The driving forces behind the smart city development in China

There are three important factors driving the development of smart cities in China: urbanization, policies, and technologies.

### 3.1 Urbanization

As a city develops, it attracts increasing numbers of immigrants. According to the World Urbanization Report by United Nations, there are 180,000 people around the world settling down in cities every day (United Nations, 2018). Since the beginning of China's economic reform and opening-up, the country has been recognized for its advances in industrialization and urbanization. Data from National Bureau of Statistics of China show that more than 15 million workers from rural areas migrate to cities each year (Ning, 2021). In 1980, at the beginning of its reform process, China's urban population accounted for only 20% of the total population, a number that reached 60% in 2020, and it is expected that 80% of its population will live in cities by 2050.

The fundamental reasons that China set this ambitious rate of urbanization are twofold. First, urbanization is one of the most important engines for modernization and economic development. China's urbanization rate is lower than that of developed countries, which usually have urbanization ratios above 70%. Second, in order to sustain its economic growth, and more importantly, in the context of Sino-U.S. trade conflict, China needs to switch its export-dependent economy to a domestic consumption-driven economy. The urbanization movement can play a significant role in boosting domestic consumption, especially in the sense that per capita income of rural areas is well below that of cities in China (Bai, et al., 2014).

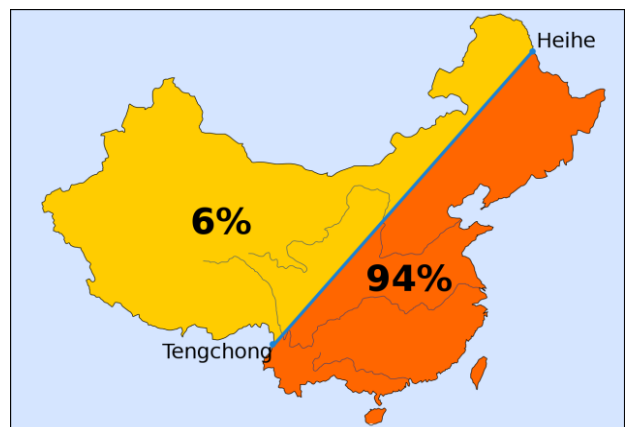

**Fig. 2  Population density on either side of the Hu Line**

As urbanization increases, the population growth occurs, which puts pressure on existing infrastructure and services. People traditionally gathered in areas conducive to their lifestyles, and made the settlement decisions based on factors such as geography, environment, history, and culture. In China, Hu Huanyong, the founder of contemporary population geography, drew the Aihui-Tengchong Line (瑷珲－腾冲线) from Aihui Heilongjiang to Tengchong Yunnan in 1935, according to population density, as shown in Fig. 2. Later, after the former city changed its name, it was known as the Heihe-Tengchong Line (黑河－腾冲线), or the Hu Huanyong Line (胡焕庸线). Today, 94% of the population live on the east side of the line, while only 6% live on the west side. In 1935, China's population was about 475 million. By 2020, it had increased to 1.4 billion, 95% of which lived to the east of the Hu Huanyong Line (Chen, et al., 2016). In the past few decades, although there have been several rounds of large migration in China (such as the reclamation of the Xinjiang Production and Construction Corps, and the development of the nation's western region), the population proportion living on either side of the Hu Huanyong Line has changed little. This means that the population in major regions and cities has increased greatly, both in terms of absolute quantity and population density.

This continuous population growth and urban density increase posed significant challenges to cities' resources and infrastructure. The traditional urban governance model is incapable of solving the additional demands for transportation, medical care, education, environment, management and other factors caused by the increased population. A new urban operational and governance model is therefore urgently needed.

### 3.2 Policies

The second important driving force for the smart city development in China is national strategic plans and policies. The concept of the smart city originated by David Bollier of University of Southern California, in his 1998 book - How Smart Growth Can Stop - Sprawl (Bollier, 1998). Bollier advocated the abandonment of the traditional city construction concept, which countered the problems of urbanization by setting up more infrastructure. Instead, he proposed that new smart city planning solutions should be adopted within limited urban spaces, which was later adopted by the city of Portland, Oregon, United States. Around 2005, Siemens in Germany and Cisco in United States also conducted research on smart cities, and when the financial crisis fulminated in 2008, IBM proposed its Smart Planet program to the Obama administration in an effort to boost the economy (McNeill, 2015). The program was intended to improve efficiency, management and quality of life for the citizens in several pilot cities, by digitizing and networking government services, and installing smart water and electricity meters to monitor the resource use and enable those cities to plan for the future. IBM also suggested using information technology to upgrade the city's infrastructure and services, such as buildings, transportation, and safety. In 2010, IBM launched the Smart Cities Challenge, which was supported by its own technology and funded at its own expense (Scuotto et al., 2016). Experts were selected to visit dozens of cities around the world to help them establish or upgrade their information technology infrastructure. As a result, the concept of the smart city gradually formed, and a construction boom swept the world. Since then, countries implem-

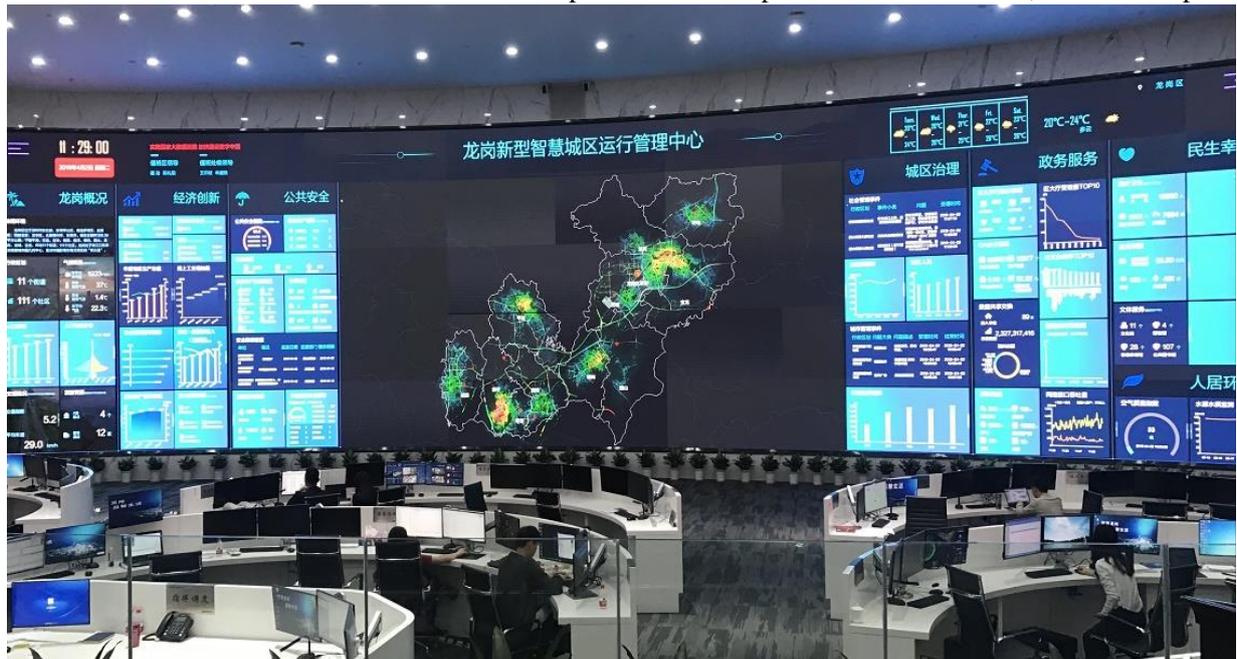

**Fig. 3 The LED screen at the Longgang Smart Center**

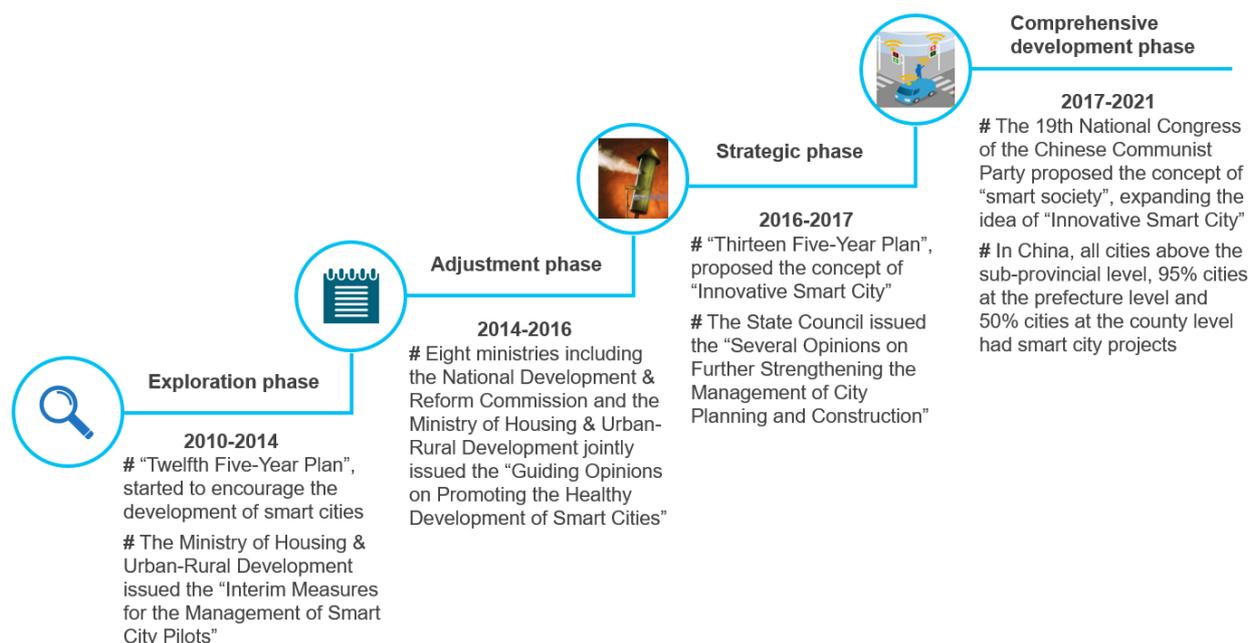

**Comprehensive development phase**

**2017-2021**
# The 19th National Congress of the Chinese Communist Party proposed the concept of "smart society", expanding the idea of "Innovative Smart City"

# In China, all cities above the sub-provincial level, 95% cities at the prefecture level and 50% cities at the county level had smart city projects

**Strategic phase**

**2016-2017**
# "Thirteen Five-Year Plan", proposed the concept of "Innovative Smart City"

# The State Council issued the "Several Opinions on Further Strengthening the Management of City Planning and Construction"

**Adjustment phase**

**2014-2016**
# Eight ministries including the National Development & Reform Commission and the Ministry of Housing & Urban-Rural Development jointly issued the "Guiding Opinions on Promoting the Healthy Development of Smart Cities"

**Exploration phase**

**2010-2014**
# "Twelfth Five-Year Plan", started to encourage the development of smart cities

# The Ministry of Housing & Urban-Rural Development issued the "Interim Measures for the Management of Smart City Pilots"

**Fig. 4 The four phases of smart city construction in China**

ented smart city strategies to transform their urban development, accelerate industrial upgrades and boost the economy.

The construction of smart cities in China can be roughly divided into four phases. The first phase took place between 2010 and 2014. It can be marked as the early stage of research and exploration. The main characteristic of this stage was the applications of information technology to digitize cities (Tang, et al., 2020). The Twelfth Five-Year Plan was released in 2010, and explicitly encouraged the development of smart cities. In 2012, the Ministry of Housing and Urban-Rural Development issued its Interim Measures for the Management of Smart City Pilots, in which the first and second batches of pilot cities (numbering 90 and 103, respectively) were selected (Kang & Wang, 2018). Longgang of Shenzhen was selected in the first batch, and a smart city operation and management center was established. The authors visited the Longgang Smart Center twice. The staff explained that the air, water, hillside slopes, fire facilities, roads, houses, streetlights, cameras, manhole covers and other infrastructure in the pilot zone were all equipped with sensors and computing units. Data collected by these sensors are processed and analyzed at the Center, which then has an overall picture of Longgang (including its joint command under emergency conditions and its big data decision-making support). All of these can be viewed on a high-quality

168 m2 LED screen, which is currently China's biggest indoor LED screen with the largest curvature and the highest resolution (Fig. 3). The city was no longer an agglomeration of cold hardware (steel bars, water pipes, roads and wires), but an intelligent body of information with an inherent sensing and computing cycle: perception, connection, computing, and intelligence.

The second phase of China's smart city development took place between 2014 and the beginning of 2016, and saw the strengthening of coordination between the country's ministries and commissions, aiming to regulate the disorderly development of smart cities in the first stage. In August 2014, eight ministries and commissions, including the National Development and Reform Commission and the Ministry of Housing and Urban-Rural Development, jointly issued the Guiding Opinions on Promoting the Healthy Development of Smart Cities. In October of the same year, 25 departments (including the National Development and Reform Commission, and the Ministries of Industry and Information Technology) established the Smart City Inter-Ministry Coordination Working Group, to coordinate regional smart city construction.

The third stage, from 2016 to 2017, was a strategic one. With the issuance of the Thirteenth Five-Year Plan at the beginning of 2016, the smart city concept started to become a national strategy. The plan was the first high-level document to mention the concept of "Innovative Smart City", which proposed the building of a group of innovative and demonstrative smart city pilots (Li, et al., 2017). Building innovative smart cities allows the "data chimney" or "data island" (in which data systems are separate, and unable to coordinate with each other) to be replaced by integrated information systems.

The fourth stage, from the end of 2017 to the present, has been a period of comprehensive development of smart cities. In December 2017, the report of the Nineteenth National Congress of the Chinese Communist Party first proposed the construction of a "smart society" (Liang, et al., 2019). This was the first time the concept, which entails the sinicization and modernization of smart cities, appeared in a national policy report. The smart society has broader implications, and is more citizen-oriented, than smart cities. In China to date, all cities at or above the sub-provincial level, 95% of those at the prefecture level, and 50% of those at the county level have all implemented smart city projects.

The four phases of smart cities in China are summarized in Fig. 4.

### 3.3 Technologies

The third important driving force for smart city is the development and maturity of information and communication technologies, which began in the last century. Internet, Internet of Things, big data, cloud computing and artificial intelligence have been widely applied in company operations, municipal administration, and daily lives of individuals. These information and communication technologies provided the essential conditions for the development of smart cities.

First, sensors, such as thermistors, photoresistors, tachometers, pressure gauges and gyroscopes, perceive subtle changes in a city's functional aspects, and give these physical changes a digital form. This digitization process enables computing devices to implement data transmission, visual presentation, and data analysis, and to engage in intelligent decision-making. More importantly, digitization enables the development of new applications, and the establishment of new business models. For example, health monitoring is a typical information-intensive application. Only when physical changes are faithfully collected by sensors and timely analyzed by algorithms, can new applications or products be developed. The taxi industry, the logistics supply chain and municipal units of administration, are typical examples as well, where digitization can link entities with different demands through the data collected by various sensors.

Second, the collected sensor data must be sent to its destination accurately and efficiently. This can be done through different transmission mediums (such as optic fibers, twisted pair cabling or electromagnetic waves), and via various network forms (personal area network - RFID, NFC, Bluetooth, local area network - WiFi, wide area network - LoRa, NB-IoT and 5G), which have already been widely operating in different business, industrial and life scenarios (Habibzadeh, et al., 2018).

Third, data arrived at the destination needs to be processed. To avoid excessive data being transferred or processed, one method is to use simple algorithms to pre-screen data at sensors, sink nodes or gateways, thus reducing the volume of information that needs to be transmitted and stored. This allows quick decisions to be made, which is usually referred as the edge or fog computing. Another method is to converge data, and process them at regional, provincial or even national computing centers, where global optimization or optimal decisions can be made based on available information from different aspects. This is known as the cloud computing.

These different levels of information technologies are used to collect, share, transmit, aggregate and compute data from all areas of a city, and help citizens or municipal administrators make informed and optimized decisions. This process plays a significant role in how new cities are governed, and how the challenges of metropolises are addressed.

The above-mentioned technological model of smart cities, sensing-networking-data analyzing, can be exemplified by Smart Shenzhen (Hu, 2019) (Yeung & Lu, 2020). The Smart Shenzhen gave a city-wide integrated view toward smart cities, which included: one map for comprehensive perception, one e-ID to travel throughout Shenzhen, one-click to know the overall situation, one integrated operation linkage, one-stop innovation and entrepreneurship, and one screen to enjoy Shenzhen (Shenzhen Government, 2018). The goals of "six one" are to enhance people's livelihood and governance capacity, which are achieved via three abstracting steps: 1) various sensors to collect city data, 2) obtained data to be transmitted via different networking technologies, and 3) data arrived at destination to be analyzed and presented to machines to make automatic decisions or to citizens in different forms of urban services, as depicted in Fig. 5.

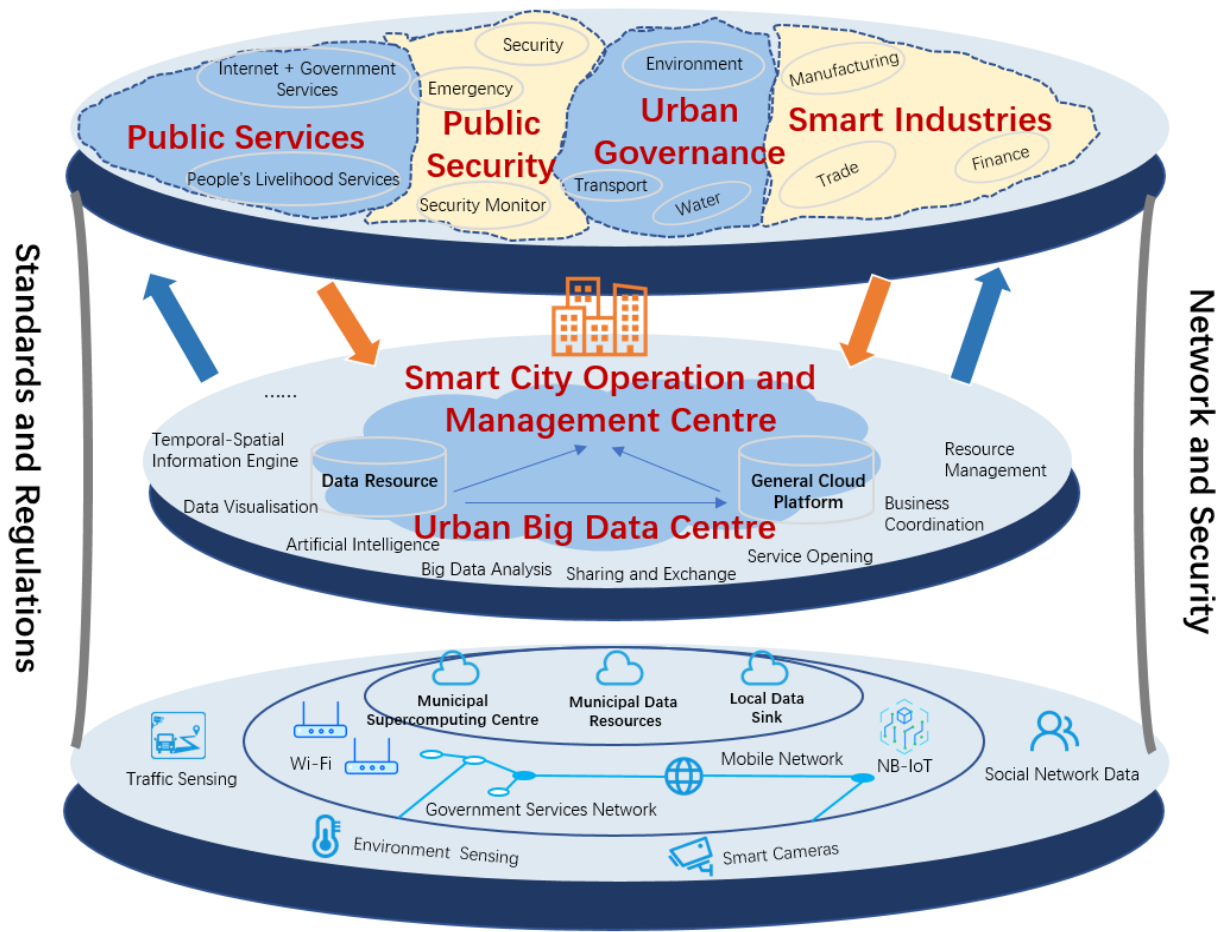

**Fig. 5 The smart city structure of Shenzhen**
**Sources: Recreated based on Shenzhen Government (Shenzhen Government, 2018) and Energies (Hu, 2019).**

## 4 Concluding remarks

The United Nations 2030 Agenda for Sustainable Development listed seventeen goals and challenges for achieving a better and more sustainable global community for all, as shown in Fig. 6 (United Nations, 2015). The goals and challenges include: eradicating poverty and hunger, making cities and communities sustainable, ensuring that populations are healthy and have access to quality education, clean water, sanitation and affordable clean energy. The top-level design of smart cities in many countries is based on the UN's 2030 Agenda.

In the above sections of this paper, we discussed what a smart city is, why a smart city is needed, and its core driving forces. In this section, we will discuss the challenges of building a smart city from four aspects: 1) the coexistence of new and old infrastructure, 2) the accessibility of city services, 3) the sustainability of smart cities in China, 4) the importance of comprehensive policies, laws and regulations, and suggestions on how these challenges can be addressed.

### 4.1 The coexistence of new and old infrastructures and functions

The first challenge in the construction of smart cities is to ensure the coexistence and integration of new and old infrastructures. Cities provide services to their citizens through infrastructures and functional units. As shown in Fig. 1, a typical characteristic of a smart city is to embed the information technology empowerment layer into the gap between infrastructures and services, thereby using a range of technolo-gies to optimize urban operations and citizens' quality of life. In the early stages of smart city development, most of these newly developed services and applications were promoted and implemented by government departments. Now, however, increasing numbers of individuals, groups and companies are using the newly-available data created by the information technology empowerment layer to develop smart applications and new business models, which link a city's functional units with its citizens. Interestingly, in most cases, it is not done by the owners of the infrastructure, but rather by visionaries and entrepreneurs, whose new applications affect traditional service providers. The ride-sharing platform Didi, for example, does not own the means of production or the infrastructure it uses. Although, it did not affect Didi's rapid rise, the platform's success did lead to strikes and complaints from traditional taxi drivers.

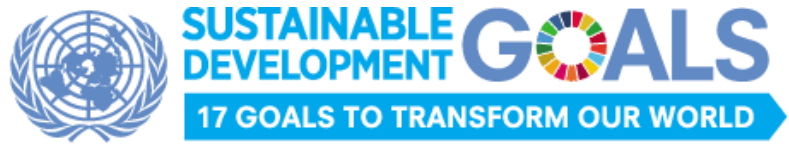

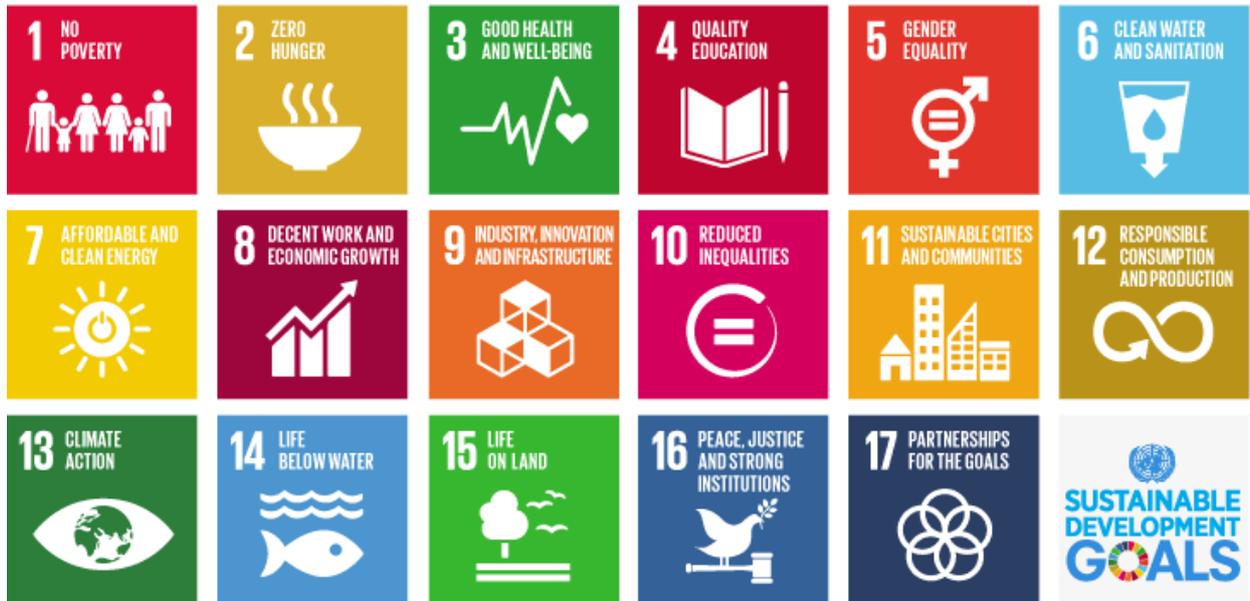

**Fig. 6 The United Nations Sustainable Development Goals**

In addition, the positive externality of urban networks leans on the providers of emerging services rather than the builders or owners of infrastructures. Although internet service providers need to build a large amount of infrastructures to provide services to citizens, most of the benefits are obtained by providers of big data services or innovative internet applications. The externality issue does not only exist in urban communication networks, but also applies to other urban infrastructures. This is problematic, because the providers of emerging services do not take charge of the construction, maintenance and development of urban infrastructure. In this regard, the first challenge in the construction of smart cities is to coordinate the harmonious coexistence of new and old facilities and functional units, as only in this way, we can ensure that new services are continuously created.

**4.2 Service accessibility**

The second challenge in the construction of smart cities is to ensure that the services they provide are accessible, in both physical and economic terms. Some traditional urban services are at the mercy of the digital economy, so they gradually lose their market shares to new business models. The new services based on advances of information technology are considered convenient and easy to use by ordinary people. However, some disadvantaged groups cannot or do not know how to use smart terminals, because of factors like age, physical status, or cultural exposure. As a result, they are excluded from the full scope of new services. Further, some new services have increased in price due to market or information monopolies. It implies that new services cater more to young, educated or wealthy people, which is pre-sumably not in line with the original intention of smart city designers. Therefore, in addition to providing new applications, smart city builders must provide alternative services that are accessible to everyone on an equal basis.

### 4.3 Sustainability of smart city projects

Compared with smart city initiatives in other countries, there are two notable features of smart cities in China: technology centrism, and top-down paradigm. Despite its national strategy of smart city emphasized 'people-oriented', China pursued constructing smart cities almost exclusively on the latest advancement of technologies, e.g., IoT, 5G, big data, AI and cloud computing (Huang, et al., 2021) (Liu, et al., 2019). In addition, the smart city initiatives in China are heavily driven by national policies set up by higher governments or ministries. The local smart city policies will not deviate from the upper-layer guidelines. This top-down paradigm is really different with the philosophy of smart cities in the West, where local governments and private entities usually take the lead (Hu, 2019) (Li & Li, 2016).

The technology-centric and top-down features of smart cities may result in the sustainability issue in the long run. Because the former feature has a narrow focus on technology and infrastructure (Liao, et al., 2016), thus lacking different stakeholders' involvement, while the latter fails in building up a direct connection and relevance to local, thus possibly making their smart city trials to be a spherical cow.

### 4.4 Improving policies, laws and regulations

The fourth challenge is to improve the legal system. In general, the formulation of policies, laws and regulations always lags behind technological innovations (Liao, 2020). This presents a big challenge to ensure that new technologies or services are not used for illegal purposes before they are properly regulated. The anonymity of blockchain technology, for example, can be used by cryptocurrency platforms for tax evasion or money laundering. However, it does not mean that we should stifle or abandon the exploration of new technologies because of the potential for illegality in certain applications. Some cities have pioneered the use of a supervision sandbox to address this legal issue. A supervision sandbox is a closed virtual testing ground, where smart city builders and application developers can confine new technologies or applications in a controlled environment for experimentation. It is usually used to develop new business models that are not covered by existing laws or regulations (Zetzsche, et al., 2017).

In addition, data is the new oil in the information age, which means money and sometimes authority. Thus, its collection, storage and ownership are challenges that smart city builders must address (Hou, et al., 2021). In July 2021, Didi was temporarily removed from app stores because its collection and use of personal information violated the Cybersecurity Law of China, which came into effect in June 2017 (National People's Congress of China, 2017). On 10 June 2021, the National People's Congress passed the Data Security Law, which came into effect on 1 September 2021, and is stricter than the Cybersecurity Law in terms of data classification, data localization and cross-border data control (National People's Congress of China, 2021). The Data Security Law effectively supplements and improves the Cybersecurity Law, because it ensures the orderly and legal flows of data. With these tools at their disposal, smart city designers and new service developers must find a balance between protecting citizens' data security and using this data to provide them with better services.

**Contributors**

Ruizhi LIAO orchestrated this research work and drafted the manuscript. Liping CHEN helped organize the manuscript. Ruizhi LIAO and Liping CHEN revised and finalized the paper.

The authors would like to thank the editors and anonymous reviewers for their careful reading of our manuscript and offering insightful comments. The authors are also grateful for the figures processed by Zhan SHI and Qianyu OU at The Chinese University of Hong Kong, Shenzhen, China.

**Compliance with ethics guidelines**

Ruizhi LIAO and Liping CHEN declare that they have no conflict of interest.